# The separation of market and price in some free competitions and its related solution to the over-application problem in the job market

Vincent Zha
ORCID: https://orcid.org/0000-0002-8435-9111

**Abstract**

According to common understanding, in free completion of a private product, market and price, the two main factors in the competition that leads to economic efficiency, always exist together. This paper, however, points out the phenomenon that in some free competitions the two factors are separated hence causing inefficiency. For one type, the market exists whereas the price is absent, i.e. free, for a product. An example of this type is the job application market where the problem of over-application commonly exists, costing recruiters much time in finding desired candidates from massive applicants, resulting in inefficiency. To solve the problem, this paper proposes a solution that the recruiters charge submission fees to the applications to make the competition complete with both factors, hence enhancing the efficiency. For the other type, the price exists whereas the market is absent for a product. An example of this type is the real estate agent market, where the price of the agents exists but the market, i.e. the facility allowing the sellers' information to be efficiently discovered, is largely absent, also causing inefficiency. In summary, the contribution of this paper consists of two aspects: one is the discovery of the possible separation of the two factors in free competitions; the other is, thanks to the discovery, a solution to the over-application problem in the job market.

## 1. Introduction

According to common understanding, in free completion of a private product, market and price, the two main factors in the competition that leads to economic efficiency, always exist together [1, 2, 3, 4]. To clarify, in this paper the term market refers to the facility allowing the sellers' information to be efficiently discovered. For example, grocery stores serve as a market allowing merchants to be easily discovered by buyers. Without the stores, it will be difficult for buyers to know and buy from the individual sellers located far away. For another example, the internet is a market for many kinds of products, letting buyers know the sellers efficiently. Also, the term price in this paper emphasizes a value greater than zero. That is, the price of zero, or free, is considered the absence of price. In most cases of free competitions, both market and price exist.

In free competition of some private products, however, market and price don't exist together. One factor can exist whereas the other is absent. The discovery of this phenomenon is a contribution of this paper. The discovery leads to solutions to some problems in the market hence enhances the economic efficiency.

## 2. Theory development

Contrary to common understanding, this paper reveals that is it possible that market and price can be separated in free competitions. To illustrate, this paper describes the four possible combinations of the statuses of the two factors in Fig 1.

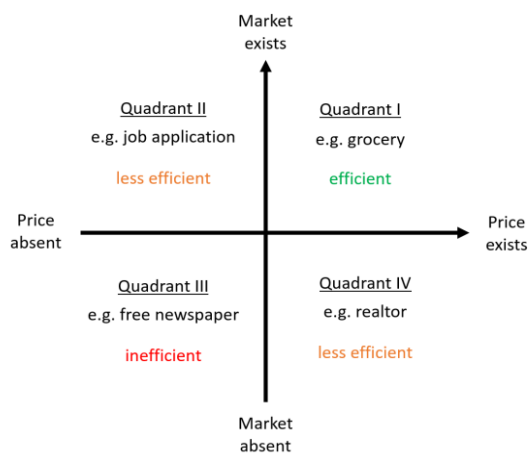

Fig 1. Combinations of market and price status in free competitions

In Fig 1, Quadrant I illustrates the situation where both market and price exist in free competitions. Most products belong to this quadrant. For example, groceries normally have both market and price, resulting to economic efficiency.

The discovery of the existence of Quadrants II and IV, however, is this paper's contribution.

Quadrant II illustrates the situation where market exist but price is absent in a free competition. An example is the job application market. Here, the "sellers" are the recruiting institutions. The "product" that they "sell" is the recruiters' time to read applications and conduct interviews. The "buyers" are the applicants. On one hand, the market exists as the job posts can be easily discovered through the internet; on the other hand, the price, however, is absent. That is, candidates submit applications normally free of charge. Therefore, the job application market belongs to this quadrant.

It often happens that massive applications are submitted to job posts, costing recruiters much time in filtering through and finding desired candidates. The over-application problem leads to economic inefficiency. This paper proposes a solution in a later section, thanks to the discovery of the quadrant.

Quadrant IV illustrates the situation where price exists but market is absent in free competition. An example is the real estate agent market. Here, the sellers are the realtors. The product is their expertise. The buyers are the house traders hiring the agents. The price, which is the realtors' commission, is normally clear, for example, 2.5% of the property price. The market, however, is largely absent. That is, the information of the agents cannot be efficiently discovered by house traders. It is common that a trader learns about and hires an agent from the trader's friend. The information source of all available realtors is very limited. Although agents make advertisements, traders often rely on recommendations by acquaintances, as agents' qualities cannot be easily discovered through advertisements. Therefore, the real estate agent market belongs to this quadrant.

Similar to the realtors' market, many other professional service markets belong to this quadrant.

Like Quadrant II, Quadrant IV is less efficient, too. In the realtor's market, the traders often miss the chance of hiring better agents due to the lack of the market. A hint of the inefficiency is the largely fixed commissions of all agents. In a highly competitive hence efficient market in Quadrant I, however, the price difference between the high and low quality products will be much larger, effectively rewarding the high quality products and punishing the low one.

Quadrant III illustrates the situation where both price and market are absent in free competition. An example is the free newspaper where the price is free and readers are largely unaware of alternative free newspapers.

**3. Theory application**

**3.1 Solution to the over-application problem in job market**

Thanks to the discovery of the separation of market and price in free competitions, this paper proposes a solution to the over-application problem in the job market. Since the problem is caused by the missing factor of price in the competition, the solution is to add a charge to applications. Therefore, the proposal makes the recruit market complete with both factors, hence moving from Quadrant II to Quadrant I, enhancing the efficiency.

To clarify, the proposal is to let the recruiters announce a price for the candidates to pay as a condition to application submission. The price will be freely decided by the recruiters. The solution will reduce the number of applications to a desired amount. In essence, the solution is no difference from a regular product, where the good quality products are priced high and the low quality ones are priced low. Similarly, in job application, a reputable hiring institution can charge a high price on submission, whereas a less competitive institution has to charge less, or even negative in some extreme conditions [5].

**3.2 How the solution works**

Once the solution is applied, many weak candidates for job posts will not submit applications as they will not want to pay for the low chance of recruitment. The solution makes the difference from the current free application market. As a basic economic rule, anything free is tend to be abused. On the other hand, the strong candidate will mostly still apply as they deem the chance of recruitment high enough to justify the submission cost.

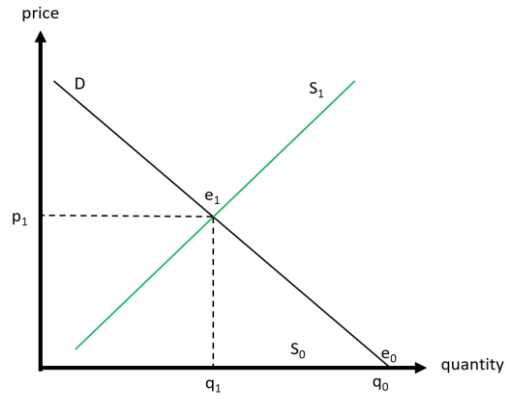

*Fig 2. effect of the solution to the over-application problem*

The effect of the solution can be illustrated by the supply and demand curves in Fig 2. The demand curve D is a normal tilted line. It does not change by the solution. The current supply curve $S_0$, however, is an uncommon flat line overlapping the x-axis, reflecting the absence of the price. Both curves intersect at point $e_0$ where the equilibrium is reached. At $e_0$, the equilibrium price is 0 and the equilibrium quantity is $q_0$, which is very high, reflecting the over-application problem due the free submission.

With the solution, the supply curve will move from $S_0$ to $S_1$, becoming a normal tilted line. It reflects the recruiters' charged services. The new intersection point $e_1$ reflects the new equilibrium, where the new equilibrium price will be $p_1$ which is greater than 0, and the new equilibrium quantity will be $q_1$, which is less than $q_0$, reflecting that the solution reduces the amount of applications.

In reality, most of the submissions prevented by the solution will be the cases where the candidates know that their qualifications largely mismatch the job requirements.

**3.3 Counterarguments to potential concerns**

As the solution is unconventional, there may be concerns on it. This paper provides counterarguments.

**3.3.1 Counterargument to the concern that the solution will hurt poor candidates**

A concern may be that, the poor candidates will not be able to pay the submission fees hence lose the chance of recruitment.

The counter argument of this paper is that, with the assumption of rational individuals, the poor candidate will still apply to matched job posts by prioritizing budget or borrowing money, as the potential reward of recruitment justifies the relatively small submission cost.

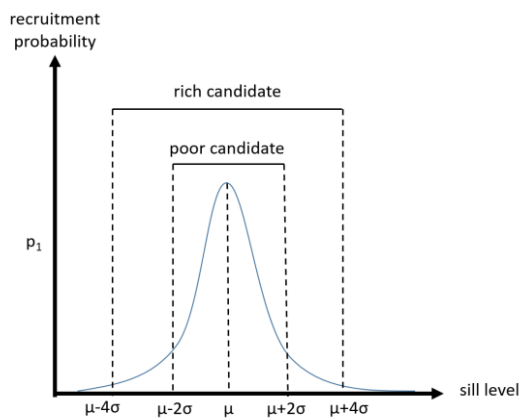

*Fig 3. different effects of the solution to the poor and rich candidates*

In fact, what the solution prevents is the abuse of application, which is illustrated in Fig 3. The requirements of all posts in the market are simplified into a 1-dimensional factor, i.e. skill level, as represented by x-axis. The y-axis is the candidate's probability of recruitment. A candidate's best chance of recruitment is at the job post matching the candidate's own skill level, which is µ. The candidate may apply to the posts requiring skill levels higher or lower than the candidate's own level, and the chance of recruitment reduces as the required skill level moving away from µ. This paper uses a normal distribution curve to reflect the situation in the figure.

A rational candidate will try to maximize the overall chance of recruitment; therefore, the candidate will apply for jobs with the required skill levels centred at µ.

From Fig 3, the different effects of a submission fee on poor and rich candidates can be seen. For a poor candidate, due to limited budget, the person will have to limit the applications to a relatively small range, say, within 2σ of µ. Here, the σ is the standard deviation of the normal distribution. Conversely, for a rich candidate, the person will try to maximize the chance by expanding the submission range to, say, 4σ, thanks to the higher budget. It can be seen from Fig 3 that the rich candidate will need to make a lot more applications than the poor candidate to cover the range of 4σ, but the gained additional recruitment chance is very small, due to the nature of the fast diminishing tails of the normal distribution curve. Put another way, what the submission fee prevents the poor candidates from doing is the massive submissions that largely have no chance of recruitment.

### 3.3.2 Counterargument to the concern that the recruiters may lose the chance of hiring the best candidates

Another concern may be that, because it is not easy for candidates to correctly evaluate their own strength, some truly matching candidates may not apply due to wrong self-assessment.

The counter argument of this paper is that, in terms of possibility, truly matching candidates will be more likely to deem themselves matching, whereas truly mismatching candidates will be more likely to deem themselves mismatching. Therefore, the matching candidates will be still likely to apply.

### 3.3.3 Counterargument to the concern of recruiters' hesitance out of worry about image hurt

Still another concern may be that, hiring institutions may be hesitant to charge a submission fee worrying about the hurt to their image.

The counter argument is that, the solution will bring efficiency hence benefit the hiring institution. Also, it commonly happens in the educational area where colleagues, as well as private high schools and primary schools, charge application fees. This phenomenon, although not fully comparable, can serve as a reference of feasibility.

A practical method to protect the recruiters' images at the beginning phase of the solution could be that, the recruiters make an announcement that they will (double) donate the application fees to the applicants' preferred charities. The hiring institutions in the job market should be eventually adapt to the solution. The efficiency enhancement will ultimately benefit the society and will be deemed as a common and reasonable practice in the future.

## 4. Conclusions

This paper reveals the phenomenon of the separation of market and price in some free competitions. Thanks to the discovery, this paper proposes a solution, which is for the recruiters to charge a submission fee to the applications, to solve the over-application problem that commonly exists in the job market. Future work includes further analysis to discover more special products and markets that comply with the phenomenon, and the attempt to find solutions to enhance their efficiency according to the discovery.

**References**

1. Joan Robinson, What is Perfect Competition? The Quarterly Journal of Economics, Volume 49, Issue 1, November 1934, Pages 104–120
2. Paul J. McNulty, A Note on the History of Perfect Competition, Journal of Political Economy, Vol. 75, No. 4, Part 1 (Aug., 1967), pp. 395-399
3. Clark, J. M. Competition as a Dynamic Process., Washington: Brookings Institution, 1961.
4. Eduardo M. Azevedo, Daniel Gottlieb, Perfect Competition in Markets With Adverse Selection, Econometrica, 2017
5. https://www.newsweek.com/mcdonalds-restaurant-offers-job-applicants-50-turn-interview-1585692